\documentclass[sigconf,nonacm]{acmart}

\AtBeginDocument{%
  \providecommand\BibTeX{{%
    \normalfont B\kern-0.5em{\scshape i\kern-0.25em b}\kern-0.8em\TeX}}}

\setcopyright{rightsretained}
\acmYear{}
\acmConference[]{Mensch und Computer 2021}{Workshopband}{Workshop on User-Centered Artificial Intelligence (UCAI '21)}
\acmISBN{}
\acmDOI{10.18420/muc2021-mci-ws02-232}

\begin{document}

\title[Audit, Don’t Explain - Recommendations for ML]{Audit, Don’t Explain – Recommendations Based on a Socio-Technical Understanding of ML-Based Systems}

\author{Hendrik Heuer}
\email{hheuer@uni-bremen.de}
\orcid{0000-0003-1919-9016}
\affiliation{%
 \institution{University of Bremen, Institute for Information Management}
 \city{Bremen}
 \country{Germany}
 }

\renewcommand{\shortauthors}{Heuer}

\begin{abstract}
In this position paper, I provide a socio-technical perspective on machine learning-based systems. I also explain why systematic audits may be preferable to explainable AI systems. I make concrete recommendations for how institutions governed by public law akin to the German TÜV and Stiftung Warentest can ensure that ML systems operate in the interest of the public.
\end{abstract}



\maketitle

\section{Motivation}

My research explores ML-based systems from a critical perspective that considers the potential advantages and examines the political, social, and individual implications of such systems. The goal of this paper and my other work \cite{elib_4444,10.1145/3415192,10.1145/3340631.3394873} is to contribute towards Suchman's goal of `lessen[ing] the asymmetry [in relative access to the contingencies of the unfolding situation] by extending the access of the machine to the actions and circumstances of the user' \cite{suchman_human_machine_2007}. As a first step towards this goal, the paper recognizes the agencies and attendant responsibilities in the context of ML-based systems, extending on Suchman \cite{suchman_human_machine_2007}.

I extend on \cite{Eslami2015} by investigating the larger socio-technical system around ML-based systems and by recognizing the role of ML practitioners, providers of data, the organization that operates the systems, and other users. In addition to that, I motivate why auditors are needed to control ML-based systems as important `public relevance algorithms' \cite{Gillespie2014}.

This paper responds to Konstan and Riedl's call that the user experience of such ML-based systems needs further attention \cite{Konstan2012}. As suggested by Jannach et al. \cite{jannach2019towards}, this paper does not aim to achieve small increases in prediction accuracy. Instead, I provide a broad understanding of users and ML-based systems and actors in the socio-technical system.

This distinction of the different actors that influence ML-based systems and their goals directly relates to Alvarado et al.'s notion of algorithmic experience and how it differs from user experience \cite{Alvarado2018}. As argued, a system like YouTube's ML-based system can have an excellent user experience and a poor algorithmic experience. However, to make this distinction, a thorough understanding of the agencies and attendant responsibilities in the context of a particular technology provided in this paper is crucial.

\section{Agencies \& Attendant Responsibilities in ML}

Based on the work in my doctoral thesis on users \& machine learning-based curation systems \cite{elib_4444}, I want to highlight different agencies in ML systems. My research indicates that it is not sufficient to merely investigate the primary users of a system, i.e., those who use YouTube or Facebook. In addition to users, data, the ML algorithm, the inferred model, and the output of the ML system, my research provides accounts on at least five other actors that influence ML-based systems: 

\begin{itemize}
\item `the organization' that operates the system
\item `other users', e.g. in a social media setting
\item `providers of data', who provide labeled data to train ML-based systems
\item `ML practitioners', who develop and evaluate the ML-based system
\item `auditors', who audit ML-based systems based on their output
\end{itemize}

It is important to examine these different actors, their responsibilities, and their goals to recognize the agencies and understand how ML-based systems can and should be explained and audited.

For instance, `the organization' will develop and operate an ML-based system for a specific goal \cite{10.1145/3415192}, e.g., providing recommendations to users. This, however, may not be the primary objective of the organization. The primary goal could be increasing revenue or driving share prices. Therefore, goal conflicts could arise when a company's primary objective influences the ML-based system and how it provides recommendations.

`Other users' potentially influence the ML-based system through their actions \cite{10.1145/3415192}. However, they may not even be aware that the ML-based system exists. In addition to that, they most likely are not aware of how their actions influence the ML-based system. For this reason, they may not realize their responsibilities and the significance of their actions. The goals of `other users' may be related to finding particular content, which may or may not align with the goal of the ML-based system as a whole. Therefore, it is important to deeply involve users \cite{jarke2021co} and to investigate their perception of systems \cite{10.1007/978-3-030-57717-9_14,10.1145/3377325.3377498}.

`Providers of data' also have a lot of responsibility for ML-based systems and the quality of the recommendations \cite{Heuer:2018:TNS:3240167.3240172}. They directly influence the quality of the data. Meanwhile, those who provide the data, e.g., through crowdsourcing, might not know what the data is actually used for. Their goal could be earning a small fee for rating some data points. Here, again, the goals of the `ML providers' may not be aligned with the goals of the ML-based system. Providers of data may not even be aware that their data is used to train an ML-based system. This is especially true in cases where existing data is repurposed to train the system, e.g. ImageNet. 

`ML practitioners' are those who train and develop ML-based systems. They are another group of people that directly influence the ML-based systems \cite{bds_paper}. Their objective is to develop and evaluate the ML-based systems as well as the interface with which users interact. This is crucial regarding Dove et al.'s finding an understanding of ML and its applications is only emerging among designers and user experience experts \cite{Dove:2017:UDI:3025453.3025739}. While the goals of `ML practitioners' can be expected to be most closely aligned with the goals of the ML-based systems they developed, the contemporary understanding of ML is still limited. Therefore, even those who train ML-based systems may not fully understand how and why a system works in the way it works \cite{knight_dark_2017}.

`Auditors' are another relevant group of actors recognized in this paper \cite{elib_4444}. Auditors systematically investigate the input-output correlations of ML-based systems, e.g., following the model by Sandvig et al. \cite{sandvig2014auditing}. Their goal is to assert accountability of ML-based systems and to ensure that the system does not enact systematic biases. My research demonstrates why audits by independent `auditors' may be preferable to explanations for the `current user' \cite{elib_4444}.

The findings in \cite{bds_paper} add to this by recognizing the role of `ML practitioners' in developing and evaluating ML-based systems. However, the investigation also revealed essential limitations regarding how the significance of data is discussed. In the paper, we examined practitioners' framing of machine learning and demonstrated why `algorithms' are not the central issue when critically reflecting on ML-based systems. Based on this, we argue that the significance of data in ML-based systems and machine learning in general needs more attention.

Recognizing `the organization' and `other users' directly relates to the user belief framework presented by Alvarado et al. \cite{10.1145/3415192}. The paper showed that even users without a background in technology recognize more than their own influence as the `current user' of a system. The user belief framework evinces that users also consider the role of the `ML algorithm' on who and what is considered similar and the role of social media and `other users'. In addition to that, even users without a background in technology exhibit an intuitive sensibility for the socio-technical nature of ML-based systems by considering the role that company policy and `the organization' have. 

\section{Audits of ML-Based systems}

Based on these insights, I make the following recommendations for the analysis of ML-based systems. Rather than developing explanation systems or legally requiring platform providers to provide explanations of ML systems, I recommend auditing ML systems by systematically investigating input-output correlations of such systems, following the model by Sandvig et al. \cite{sandvig2014auditing}. Scraping audits and sock puppet audits are, in my opinion, the most promising method to investigate complex ML systems.

Considering the limitations of explanations \cite{elib_4444,10.1145/3340631.3394873}, this paper recommends adopting algorithm audits to enable individuals and collectives to ensure that ML-based systems act in the public's interest. One criterion could be ensuring that all different political opinions are given sufficient room for expression. Audits could also be used to determine whether a system is enacting a gender bias or if the system has a tendency to discriminate against or towards a particular ethnic group. Controversial political topics require a balanced presentation of all arguments. This is not just a normative statement. It is required by law in Germany, where private broadcasting services must generally reflect a plurality of opinion \cite{interstate_broadcasting_agreement}.

By investigating input-output correlations, algorithm audits can be conducted independently of the platform provider \cite{elib_4444}. This enables researchers, non-governmental organizations, lawmakers, and other stakeholders to understand predictions by complex ML systems that would be hard to investigate otherwise. This could enable stakeholders to scrutinize the actions of such ML-based systems and to punish offenses, e.g., if such systems do not comply with current and future laws. This, in my opinion, would be the most important and immediate step that needs to be taken for public relevance algorithms like YouTube's recommender system and Facebook's News Feed algorithm.

In the long term, institutions need to be established to enforce laws like the \textit{Rundfunkstaatsvertrag} (Interstate Broadcasting Agreement) \cite{interstate_broadcasting_agreement} and to monitor the activity of ML-based systems on platforms like YouTube. Such institutions should be governed by public law, i.e., they should be independent and reliably financed. The public control of ML could follow the model of the German Association for Technical Inspection (Technischer Überwachungsverein - TÜV). The services of the German TÜV are required in a variety of contexts. TÜV institutions, for instance, evaluate each car in Germany every second year to ensure that the car is street-legal. A related model is the German Foundation for Product Testing~(Stiftung Warentest), akin to the Consumers Union in the U.S. and the Union Fédérale des Consommateurs in France \cite{wiki_warentest_2019}. The purpose of the German Foundation for Product Testing is to compare goods and services in an unbiased way. 

The German TÜV ensures that something complies with a certain norm -- commonly making binary decisions whether something is permitted or not. The Stiftung Warentest usually develops a catalog of criteria used to compare different instances of a specific kind of product or service. An expert consortium defines these criteria for specific products or services and a particular context. A Foundation for ML-based Systems could adopt this schema and iteratively develop criteria for the control of ML-based systems.

I hope that this paper will inspire other researchers to examine users' understanding of ML-based systems and motivate them to design and develop novel ways of explaining and auditing such systems.

\begin{acks}
The work of Hendrik Heuer was funded by the German Research Council (DFG) under project number 374666841, SFB 1342.
\end{acks}

\bibliographystyle{ACM-Reference-Format}
\bibliography{references}


\begin{thebibliography}{19}


\ifx \showCODEN    \undefined \def \showCODEN     #1{\unskip}     \fi
\ifx \showDOI      \undefined \def \showDOI       #1{#1}\fi
\ifx \showISBNx    \undefined \def \showISBNx     #1{\unskip}     \fi
\ifx \showISBNxiii \undefined \def \showISBNxiii  #1{\unskip}     \fi
\ifx \showISSN     \undefined \def \showISSN      #1{\unskip}     \fi
\ifx \showLCCN     \undefined \def \showLCCN      #1{\unskip}     \fi
\ifx \shownote     \undefined \def \shownote      #1{#1}          \fi
\ifx \showarticletitle \undefined \def \showarticletitle #1{#1}   \fi
\ifx \showURL      \undefined \def \showURL       {\relax}        \fi
\providecommand\bibfield[2]{#2}
\providecommand\bibinfo[2]{#2}
\providecommand\natexlab[1]{#1}
\providecommand\showeprint[2][]{arXiv:#2}

\bibitem[\protect\citeauthoryear{Alvarado, Heuer, Vanden~Abeele, Breiter, and
  Verbert}{Alvarado et~al\mbox{.}}{2020}]%
        {10.1145/3415192}
\bibfield{author}{\bibinfo{person}{Oscar Alvarado}, \bibinfo{person}{Hendrik
  Heuer}, \bibinfo{person}{Vero Vanden~Abeele}, \bibinfo{person}{Andreas
  Breiter}, {and} \bibinfo{person}{Katrien Verbert}.}
  \bibinfo{year}{2020}\natexlab{}.
\newblock \showarticletitle{Middle-Aged Video Consumers' Beliefs About
  Algorithmic Recommendations on YouTube}.
\newblock \bibinfo{journal}{\emph{Proc. ACM Hum.-Comput. Interact.}}
  \bibinfo{volume}{4}, \bibinfo{number}{CSCW2}, Article
  \bibinfo{articleno}{121} (\bibinfo{date}{Oct.} \bibinfo{year}{2020}),
  \bibinfo{numpages}{24}~pages.
\newblock
\urldef\tempurl%
\url{https://doi.org/10.1145/3415192}
\showDOI{\tempurl}


\bibitem[\protect\citeauthoryear{Alvarado and Waern}{Alvarado and
  Waern}{2018}]%
        {Alvarado2018}
\bibfield{author}{\bibinfo{person}{Oscar Alvarado} {and}
  \bibinfo{person}{Annika Waern}.} \bibinfo{year}{2018}\natexlab{}.
\newblock \showarticletitle{{Towards Algorithmic Experience}}. In
  \bibinfo{booktitle}{\emph{Proceedings of the 2018 CHI Conference on Human
  Factors in Computing Systems - CHI '18}}. \bibinfo{publisher}{ACM Press},
  \bibinfo{address}{Montreal, Canada}, \bibinfo{pages}{1--9}.
\newblock
\showISBNx{9781450356206}
\urldef\tempurl%
\url{https://doi.org/10.1145/3173574.3173860}
\showDOI{\tempurl}


\bibitem[\protect\citeauthoryear{Bu\c{c}inca, Lin, Gajos, and
  Glassman}{Bu\c{c}inca et~al\mbox{.}}{2020}]%
        {10.1145/3377325.3377498}
\bibfield{author}{\bibinfo{person}{Zana Bu\c{c}inca}, \bibinfo{person}{Phoebe
  Lin}, \bibinfo{person}{Krzysztof~Z. Gajos}, {and} \bibinfo{person}{Elena~L.
  Glassman}.} \bibinfo{year}{2020}\natexlab{}.
\newblock \showarticletitle{Proxy Tasks and Subjective Measures Can Be
  Misleading in Evaluating Explainable AI Systems}. In
  \bibinfo{booktitle}{\emph{Proceedings of the 25th International Conference on
  Intelligent User Interfaces}} (Cagliari, Italy) \emph{(\bibinfo{series}{IUI
  ’20})}. \bibinfo{publisher}{Association for Computing Machinery},
  \bibinfo{address}{New York, NY, USA}, \bibinfo{pages}{454–464}.
\newblock
\showISBNx{9781450371186}
\urldef\tempurl%
\url{https://doi.org/10.1145/3377325.3377498}
\showDOI{\tempurl}


\bibitem[\protect\citeauthoryear{Dove, Halskov, Forlizzi, and Zimmerman}{Dove
  et~al\mbox{.}}{2017}]%
        {Dove:2017:UDI:3025453.3025739}
\bibfield{author}{\bibinfo{person}{Graham Dove}, \bibinfo{person}{Kim Halskov},
  \bibinfo{person}{Jodi Forlizzi}, {and} \bibinfo{person}{John Zimmerman}.}
  \bibinfo{year}{2017}\natexlab{}.
\newblock \showarticletitle{UX Design Innovation: Challenges for Working with
  Machine Learning As a Design Material}. In
  \bibinfo{booktitle}{\emph{Proceedings of the 2017 CHI Conference on Human
  Factors in Computing Systems}} (Denver, Colorado, USA)
  \emph{(\bibinfo{series}{CHI '17})}. \bibinfo{publisher}{ACM},
  \bibinfo{address}{New York, NY, USA}, \bibinfo{pages}{278--288}.
\newblock
\showISBNx{978-1-4503-4655-9}
\urldef\tempurl%
\url{https://doi.org/10.1145/3025453.3025739}
\showDOI{\tempurl}


\bibitem[\protect\citeauthoryear{Eslami, Rickman, Vaccaro, Aleyasen, Vuong,
  Karahalios, Hamilton, and Sandvig}{Eslami et~al\mbox{.}}{2015}]%
        {Eslami2015}
\bibfield{author}{\bibinfo{person}{Motahhare Eslami}, \bibinfo{person}{Aimee
  Rickman}, \bibinfo{person}{Kristen Vaccaro}, \bibinfo{person}{Amirhossein
  Aleyasen}, \bibinfo{person}{Andy Vuong}, \bibinfo{person}{Karrie Karahalios},
  \bibinfo{person}{Kevin Hamilton}, {and} \bibinfo{person}{Christian Sandvig}.}
  \bibinfo{year}{2015}\natexlab{}.
\newblock \showarticletitle{"I Always Assumed That I Wasn't Really That Close
  to [Her]": Reasoning about Invisible Algorithms in News Feeds}. In
  \bibinfo{booktitle}{\emph{Proceedings of the 33rd Annual ACM Conference on
  Human Factors in Computing Systems}}. \bibinfo{publisher}{Association for
  Computing Machinery}, \bibinfo{address}{New York, NY, USA},
  \bibinfo{pages}{153–162}.
\newblock
\showISBNx{9781450331456}
\urldef\tempurl%
\url{https://doi.org/10.1145/2702123.2702556}
\showURL{%
\tempurl}


\bibitem[\protect\citeauthoryear{{Federal Republic of Germany}}{{Federal
  Republic of Germany}}{2016}]%
        {interstate_broadcasting_agreement}
\bibfield{author}{\bibinfo{person}{{Federal Republic of Germany}}.}
  \bibinfo{year}{2016}\natexlab{}.
\newblock \bibinfo{title}{Interstate {Broadcasting} {Agreement}
  ({Rundfunkstaatsvertrag})}.
\newblock
\newblock
\urldef\tempurl%
\url{https://germanlawarchive.iuscomp.org/?p=655}
\showURL{%
\tempurl}


\bibitem[\protect\citeauthoryear{Gillespie}{Gillespie}{2014}]%
        {Gillespie2014}
\bibfield{author}{\bibinfo{person}{Tarleton Gillespie}.}
  \bibinfo{year}{2014}\natexlab{}.
\newblock \showarticletitle{{The relevance of algorithms}}.
\newblock \bibinfo{journal}{\emph{Media Technologies: Essays on Communication,
  Materiality, and Society}} (\bibinfo{year}{2014}), \bibinfo{pages}{167--194}.
\newblock
\showISBNx{0262319470}
\showISSN{0262319470}
\urldef\tempurl%
\url{https://doi.org/10.7551/mitpress/9780262525374.003.0009}
\showDOI{\tempurl}
\showeprint[arxiv]{arXiv:1011.1669v3}


\bibitem[\protect\citeauthoryear{Heuer}{Heuer}{2020}]%
        {elib_4444}
\bibfield{author}{\bibinfo{person}{Hendrik Heuer}.}
  \bibinfo{year}{2020}\natexlab{}.
\newblock \emph{\bibinfo{title}{Users \& Machine Learning-based Curation
  Systems}}.
\newblock \bibinfo{thesistype}{Ph.D. Dissertation}. \bibinfo{school}{University
  of Bremen}.
\newblock
\urldef\tempurl%
\url{https://doi.org/10.26092/elib/241}
\showDOI{\tempurl}


\bibitem[\protect\citeauthoryear{Heuer and Breiter}{Heuer and Breiter}{2018}]%
        {Heuer:2018:TNS:3240167.3240172}
\bibfield{author}{\bibinfo{person}{Hendrik Heuer} {and}
  \bibinfo{person}{Andreas Breiter}.} \bibinfo{year}{2018}\natexlab{}.
\newblock \showarticletitle{Trust in News on Social Media}. In
  \bibinfo{booktitle}{\emph{Proceedings of the 10th Nordic Conference on
  Human-Computer Interaction}} (Oslo, Norway) \emph{(\bibinfo{series}{NordiCHI
  '18})}. \bibinfo{publisher}{ACM}, \bibinfo{address}{New York, NY, USA},
  \bibinfo{pages}{137--147}.
\newblock
\showISBNx{978-1-4503-6437-9}
\urldef\tempurl%
\url{https://doi.org/10.1145/3240167.3240172}
\showDOI{\tempurl}


\bibitem[\protect\citeauthoryear{Heuer and Breiter}{Heuer and Breiter}{2020}]%
        {10.1145/3340631.3394873}
\bibfield{author}{\bibinfo{person}{Hendrik Heuer} {and}
  \bibinfo{person}{Andreas Breiter}.} \bibinfo{year}{2020}\natexlab{}.
\newblock \showarticletitle{More Than Accuracy: Towards Trustworthy Machine
  Learning Interfaces for Object Recognition}. In
  \bibinfo{booktitle}{\emph{Proceedings of the 28th ACM Conference on User
  Modeling, Adaptation and Personalization}} (Genoa, Italy)
  \emph{(\bibinfo{series}{UMAP '20})}. \bibinfo{publisher}{Association for
  Computing Machinery}, \bibinfo{address}{New York, NY, USA},
  \bibinfo{pages}{298–302}.
\newblock
\showISBNx{9781450368612}
\urldef\tempurl%
\url{https://doi.org/10.1145/3340631.3394873}
\showDOI{\tempurl}


\bibitem[\protect\citeauthoryear{Heuer, Jarke, and Breiter}{Heuer
  et~al\mbox{.}}{2021}]%
        {bds_paper}
\bibfield{author}{\bibinfo{person}{Hendrik Heuer}, \bibinfo{person}{Juliane
  Jarke}, {and} \bibinfo{person}{Andreas Breiter}.}
  \bibinfo{year}{2021}\natexlab{}.
\newblock \showarticletitle{Machine learning in tutorials – Universal
  applicability, underinformed application, and other misconceptions}.
\newblock \bibinfo{journal}{\emph{Big Data \& Society}} \bibinfo{volume}{8},
  \bibinfo{number}{1} (\bibinfo{year}{2021}),
  \bibinfo{pages}{20539517211017593}.
\newblock
\urldef\tempurl%
\url{https://doi.org/10.1177/20539517211017593}
\showDOI{\tempurl}


\bibitem[\protect\citeauthoryear{Jannach, Shalom, and Konstan}{Jannach
  et~al\mbox{.}}{2019}]%
        {jannach2019towards}
\bibfield{author}{\bibinfo{person}{Dietmar Jannach}, \bibinfo{person}{Oren~Sar
  Shalom}, {and} \bibinfo{person}{Joseph~A Konstan}.}
  \bibinfo{year}{2019}\natexlab{}.
\newblock \showarticletitle{Towards more impactful recommender systems
  research}. In \bibinfo{booktitle}{\emph{Proceedings of the ACM RecSys
  Workshop on the Impact of Recommender Systems (ImpactRS’19)}}.
\newblock


\bibitem[\protect\citeauthoryear{Jarke}{Jarke}{2021}]%
        {jarke2021co}
\bibfield{author}{\bibinfo{person}{Juliane Jarke}.}
  \bibinfo{year}{2021}\natexlab{}.
\newblock \bibinfo{booktitle}{\emph{Co-creating Digital Public Services for an
  Ageing Society: Evidence for User-centric Design}}.
\newblock \bibinfo{publisher}{Springer Nature}.
\newblock


\bibitem[\protect\citeauthoryear{Knight}{Knight}{2017}]%
        {knight_dark_2017}
\bibfield{author}{\bibinfo{person}{Will Knight}.}
  \bibinfo{year}{2017}\natexlab{}.
\newblock \bibinfo{title}{The {Dark} {Secret} at the {Heart} of {AI}}.
\newblock
\newblock
\urldef\tempurl%
\url{https://www.technologyreview.com/2017/04/11/5113/the-dark-secret-at-the-heart-of-ai/}
\showURL{%
\tempurl}


\bibitem[\protect\citeauthoryear{Konstan and Riedl}{Konstan and Riedl}{2012}]%
        {Konstan2012}
\bibfield{author}{\bibinfo{person}{Joseph~A. Konstan} {and}
  \bibinfo{person}{John Riedl}.} \bibinfo{year}{2012}\natexlab{}.
\newblock \showarticletitle{Recommender systems: from algorithms to user
  experience}.
\newblock \bibinfo{journal}{\emph{User Modeling and User-Adapted Interaction}}
  \bibinfo{volume}{22}, \bibinfo{number}{1} (\bibinfo{date}{01 Apr}
  \bibinfo{year}{2012}), \bibinfo{pages}{101--123}.
\newblock
\showISSN{1573-1391}
\urldef\tempurl%
\url{https://doi.org/10.1007/s11257-011-9112-x}
\showDOI{\tempurl}


\bibitem[\protect\citeauthoryear{Krieter, Viertel, and Breiter}{Krieter
  et~al\mbox{.}}{2020}]%
        {10.1007/978-3-030-57717-9_14}
\bibfield{author}{\bibinfo{person}{Philipp Krieter}, \bibinfo{person}{Michael
  Viertel}, {and} \bibinfo{person}{Andreas Breiter}.}
  \bibinfo{year}{2020}\natexlab{}.
\newblock \showarticletitle{We Know What You Did Last Semester: Learners'
  Perspectives on Screen Recordings as a Long-Term Data Source for Learning
  Analytics}. In \bibinfo{booktitle}{\emph{Addressing Global Challenges and
  Quality Education}}, \bibfield{editor}{\bibinfo{person}{Carlos Alario-Hoyos},
  \bibinfo{person}{Mar{\'i}a~Jes{\'u}s Rodr{\'i}guez-Triana},
  \bibinfo{person}{Maren Scheffel}, \bibinfo{person}{Inmaculada
  Arnedillo-S{\'a}nchez}, {and} \bibinfo{person}{Sebastian~Maximilian
  Dennerlein}} (Eds.). \bibinfo{publisher}{Springer International Publishing},
  \bibinfo{address}{Cham}, \bibinfo{pages}{187--199}.
\newblock
\showISBNx{978-3-030-57717-9}


\bibitem[\protect\citeauthoryear{Sandvig, Hamilton, Karahalios, and
  Langbort}{Sandvig et~al\mbox{.}}{2014}]%
        {sandvig2014auditing}
\bibfield{author}{\bibinfo{person}{Christian Sandvig}, \bibinfo{person}{Kevin
  Hamilton}, \bibinfo{person}{Karrie Karahalios}, {and} \bibinfo{person}{Cedric
  Langbort}.} \bibinfo{year}{2014}\natexlab{}.
\newblock \showarticletitle{Auditing algorithms: Research methods for detecting
  discrimination on internet platforms}.
\newblock \bibinfo{journal}{\emph{Data and discrimination: converting critical
  concerns into productive inquiry}}  \bibinfo{volume}{22}
  (\bibinfo{year}{2014}).
\newblock


\bibitem[\protect\citeauthoryear{Suchman}{Suchman}{2007}]%
        {suchman_human_machine_2007}
\bibfield{author}{\bibinfo{person}{Lucy Suchman}.}
  \bibinfo{year}{2007}\natexlab{}.
\newblock \bibinfo{booktitle}{\emph{Human-machine reconfigurations: {Plans} and
  situated actions}}.
\newblock \bibinfo{publisher}{Cambridge University Press}.
\newblock


\bibitem[\protect\citeauthoryear{{Wikipedia contributors}}{{Wikipedia
  contributors}}{2018}]%
        {wiki_warentest_2019}
\bibfield{author}{\bibinfo{person}{{Wikipedia contributors}}.}
  \bibinfo{year}{2018}\natexlab{}.
\newblock \bibinfo{title}{Stiftung Warentest --- {Wikipedia}{,} The Free
  Encyclopedia}.
\newblock
\newblock
\urldef\tempurl%
\url{https://en.wikipedia.org/w/index.php?title=Stiftung_Warentest&oldid=870261083}
\showURL{%
\tempurl}
\newblock
\shownote{[Online; accessed 13-December-2019].}


\end{thebibliography}

\end{document}